# Silicon Oxide Passivation of Single-Crystalline CVD Diamond Evaluated by the Time-of-Flight Technique

Kiran Kumar Kovi[*], Saman Majdi, Markus Gabrysch and Jan Isberg[*]
Division of Electricity, Department of Engineering Sciences, Uppsala University, Box 534, SE-751 21, Uppsala, Sweden.

**Abstract**

The excellent material properties of diamond make it highly desirable for many extreme electronic applications that are out of reach of conventional electronic materials. For commercial diamond devices to become a reality, it is necessary to have an effective surface passivation since the passivation determines the ability of the device to withstand high surface electric fields. In this paper we present data from lateral Time-of-Flight studies on $SiO_2$-passivated intrinsic single-crystalline CVD diamond. The $SiO_2$ films were deposited using three different techniques. The influence of the passivation on hole transport was studied, which resulted in the increase of hole mobilities. The results from the three different passivations are compared.

**Introduction**

The attractive material properties of diamond such as: wide band gap (5.47 eV), high carrier mobilities[1], high thermal conductivity, high breakdown voltage and fields, etc., make diamond a very interesting semiconductor material for high power and high frequency applications. The recent progress in the growth of high purity single-crystalline chemical vapor deposited (SC-CVD) diamond films, has resulted in increased attention and improvements towards producing diamond based electronic devices for several applications such as: MESFETs,[2] H terminated FETs,[3] p-i-n diodes[4] and high voltage Schottky diodes[5] For many electronic applications an effective surface passivation is required for better reproducibility, lower surface leakage, and lower concentration of surface trapping centers, thereby improving charge transport properties. A high concentration of surface or interface traps is normally detrimental to device functionality as unwanted charge is trapped at the dielectric/semiconductor interface and thereby affects the charge transport.

Silicon oxide and silicon nitride deposited on intrinsic thin film diamond as different passivation layers has been reported.[6] The chosen method of $SiO_2$ deposition strongly affects the oxide properties. Therefore, to achieve better passivation of diamond, detailed studies are required on $SiO_2$ deposition by different methods. In this paper, we have studied the influence of the surface passivation on hole transport in intrinsic diamond by depositing $SiO_2$ using low pressure chemical vapour deposition (LPCVD) using tetraethyl orthosilicate (TEOS), physical vapour deposition (PVD) by e-beam evaporation and plasma enhanced chemical vapour deposition (PECVD) methods.

For studying charge transport in intrinsic diamond samples, which are highly resistive, one of the most powerful methods known as the Time-of-Flight (ToF) technique, also called the transient current technique is used. It can be used both in vertical[1,8-18] or lateral configurations [19-21] to measure charge trapping, electron and hole mobilities, saturation and drift velocities. In this technique, electron-hole pairs can be created by α-particles [8,16], pulsed electron beams[22] , pulsed particle beams[23], pulsed X-rays[24] or by Q-switched UV lasers [1]. In the present study, the lateral ToF configuration was chosen to study hole transport in thin intrinsic diamond layers directly below the diamond/dielectric interface. The aim is to use this method in the future to evaluate passivation on delta-doped structures intended for field effect transistor devices where the delta doped layer is buried between the intrinsic layers.[25] The study of transport phenomena in semiconductors by the ToF technique was first presented by Haynes and Shockley [26]. An important measurable quantity is the time of transit τ, which is the time taken by the charge carriers to travel across the sample in a defined region under the influence of a known electric field. To determine the mobility in the low injection regime, bias voltage pulses *U*, synchronized with the illumination source, are applied between the contacts for a short time in order to keep the sample polarization effects at a minimum. If the illumination is focused to a region in close proximity to one of the contacts the transit of either electrons or holes can be studied, depending on the polarity of the applied bias. In case of negligible space charges, the applied electric field is given by $|\vec{E}| = |U|/d$ with *d* being the contact spacing, and the charge carriers have a constant drift velocity $v_d$. The full width half maximum (FWHM) time interval of the obtained ToF curve gives the time-of-flight τ of the electrons or holes, is given by

$$\tau = \frac{d}{\mu |\vec{E}|} + \frac{d}{v_{sat}}$$

where μ, the mobility of the charge carriers, $v_{sat}$ the saturation velocity is obtained from plotting τ versus the inverse of applied voltage as described in [21]. The results are useful in applications where $SiO_2$ is deposited as passivation layer on intrinsic as well as intentionally doped diamond films.

## Experimental

To avoid any uncertainty resulting from different transport properties of the diamond substrate all processing was done on the same sample with careful stripping of the deposited layers in between experiments to preserve the same surface roughness. One SC-CVD diamond sample with an intrinsic layer of high purity and 100 μm thickness was chosen for this study. The sample was grown on a Ib substrate using microwave plasma CVD process by *Element Six* Ltd. Prior to the deposition of $SiO_2$, the diamond sample was cleaned in a boiling mixture with equal parts of nitric acid, sulfuric acid and perchloric acid at 180-200$^0$C, followed by treatment in oxygen plasma. $SiO_2$ was deposited using the following methods:
• *LPCVD-TEOS* - The low pressure CVD process using TEOS ($Si(OC_2H_5)_4$) as a precursor is a high temperature oxide deposition process, with an operating temperature of 710 $^0$C at 0.8 Torr. The gas flows are 16 sccm TEOS, 4 sccm $O_2$ and 105 sccm $N_2$. TEOS is vaporized from a liquid source. The $O_2$ is present for neutralizing organic and organosilicon compounds. The $N_2$ dilution is used to bring the process into the right pressure regime. The rate of deposition was 4 nm/min. A film thickness of 360 nm was measured by interferometry.
• *PVD* - The $SiO_2$ was deposited by e-beam evaporation of quartz crystal, where an electron beam hits the target to transform to gaseous phase and this precipitates in solid form over the substrate, under high vacuum conditions. A film of 125 nm thickness was measured by interferometry.
• *PECVD* - $SiO_2$ is deposited by PECVD at an operating temperature of 350 $^0$C by the chemical reaction of silane and nitrous oxide in the presence of plasma. The deposition rate was nearly 50 nm/min. The thickness of the $SiO_2$ layer measured by interferometry was 380 nm.

Buffered HF was used to etch $SiO_2$ with photoresist as the mask, following patterning by standard photolithography. The etch rates for the three different oxides were: 6, 9 and 12 nm/s for LPCVD, PECVD and PVD oxides, respectively. Aluminum contacts were deposited by sputtering followed by a lift-off process. The geometry and cross-sections of the structures are shown in Fig.1. The contacts are 1.5 mm long and 0.5 mm wide. The distance between the contacts is 0.3 mm. For comparison, the same sample without any surface passivation was also prepared with the same contact geometry.

The lateral-ToF (L-ToF) configuration is used in the current study. The L-ToF set up is shown in Fig.1. A quintupled Nd-YAG laser producing short (3ns FWHM) UV pulses of 213 nm wavelength, with a repetition rate of 10 Hz is used. The beam passes through a cylindrical lens in conjunction with reflective optics, which can be adjusted to create a line focus on the sample surface. The 213 nm wavelength UV photons correspond to energy slightly greater than the band gap energy of diamond (5.47 eV). Therefore, electron-hole pairs are generated near the dielectric/diamond interface upon illumination. The line focus is a few millimeters in length, but only a few micrometers in width. The illuminated line is chosen to be in parallel and in close proximity to one of the contacts to observe carrier transport from one contact to the other.. The measurements are carried out for several bias voltages from 4V to 150V. The current measured is fed to the digital sampling oscilloscope (DSO) through the low noise amplifier, which is placed in close proximity to the sample. A more detailed description of the setup and measurement processes is given in [21].

## Results and discussion

SiO3 deposited by LPCVD-TEOS obtained good quality films with good step coverage and yields a uniform thickness all over the substrate. The deposition proceeds by diffusion, where $SiO_2$ diffuses onto the substrate uniformly. However, it is a high temperature process and has a lower deposition rate compared to other methods. With PVD (e-beam) deposition of $SiO_2$, flaking was observed due to excessive strain if the oxide films were thick. In addition, cracks were observed in the oxide layer for films thicker than 200 nm, after the sample was removed from the evaporator. For these reasons, the $SiO_2$ thickness deposited by PVD was limited to 125 nm. However, very slight cracks were still observed which affects the passivation properties. The PECVD method is a relatively low temperature process resulting in a reasonable uniformity and step coverage. The etching rates for the oxide layers varied depending on the chosen deposition method. The LPCVD deposited oxide had the lowest etch rate while the PVD deposited oxide had the highest etch rate. The etch rate can be taken as a rough indicator of the quality of the oxide, with lower etch rates implying a better (denser) oxide. Atomic force microscopy (AFM) images for the three different oxide films in an area of $2 \times 2$ μm$^2$ are shown in Fig.2. An apparent grain morphology and surface roughness of the oxide films can be observed. The RMS surface roughness obtained were 2 nm, 5 nm, 12 nm for the LPCVD, PECVD and PVD oxides respectively. An abrupt change in the grain size and poor coverage can also be observed for PVD oxide.

In the ToF measurements the illumination is strongly attenuated so as to enable measurement in the space-charge-free regime, i.e. the amount of injected charge is so small that it does not appreciably disturb the applied electric field[27]. The intensity of the illumination is monitored with a photodiode to ensure that all samples receive the same amount of illumination at the sample surface. There may still be some difference in the carrier generation rate due to variations in the reflectivity between the passivation layers, but these variations are relatively small. The oxide layers have to be transparent for the UV photons to enable the creation of electron-hole pairs in the diamond. .$SiO_2$ is transparent to the 213 nm wavelength UV photons. Current traces at different applied bias voltages were obtained from the L-ToF measurements performed for different passivation layers of $SiO_2$. Fig. 3 shows examples of current traces, obtained at bias voltages of 16, 44 and 90 volts. A prominent difference is observed in signal strengths from the graphs. Fig.3a shows the current without passivation layer, Fig.3b, c and d with passivation by LPCVD, PVD and PECVD, respectively. Transport in diamond with SiO$_2$ layers deposited by LPCVD and PECVD resulted in current traces that are roughly square as the bias voltage increases, which indicates reduced trapping at the interface. In addition to this, more charge is collected. The LPCVD sample yields the highest signal strength of all. Transport with SiO$_2$ deposited by PVD exhibits low charge collection and the full hole transit cannot be observed in the traces. This is due to excessive charge trapping resulting in a large resident space charge at the passivation/diamond interface. The observed space charge gives rise to an additional electric field that interferes with the hole transport between the contacts. Moreover, an increase in surface scattering due to poor adhesion of this layer may also affect the hole transport. The oxide deposited by PVD method (e-beam evaporation) was of poor quality both electrically and mechanically compared to the other methods in the current study. The hole mobilities for the three passivations were calculated and compared to a measurement taken without the passivation layer. The mobility for the PVD oxide case could not be calculated as the full transit could not be observed. The hole mobility was calculated from the ToF data for all the bias voltages from 4 to 150 V. Without any passivation the obtained mobility was 1300 ± 20 cm$^2$/Vs. However, with passivation the mobility increased to: PECVD: 1360 ± 25 cm$^2$/Vs and LPCVD: 1410 ± 30 cm$^2$/Vs. The increase in the mobility values can be attributed to a reduction in interface scattering and a reduced charge trapping at the diamond/passivation interface. The surface roughness and the calculated hole mobilities are given in Table 1 for comparison.

## Conclusions

It has been shown that the selection of the deposition technique is of the utmost importance when depositing SiO$_2$ passivation layers on diamond. Deposition by LPCVD-TEOS and PECVD both resulted in smooth films with good coverage. PVD (e-beam evaporation) on the other hand resulted in poor films. The influence of the chosen passivation layer on the charge transport near the interface was studied using the L-ToF technique, demonstrating the influence of surface passivation layers and the method of deposition on the charge transport in intrinsic diamond. The measured hole mobilities in intrinsic diamond varied with the chosen deposition method for the passivation layer. The highest mobility was observed using LPCVD-TEOS: 1410 ± 30 cm$^2$/Vs, which was attributed to a reduction in interface scattering and a reduced charge trapping at the diamond/passivation interface. However, the LPCVD method is a high temperature process, which may not be suitable for deposition of a passivation layer in later stages of device processing. In this case the, PECVD method can be a good alternative as it does not use as high temperatures as LPCVD, and in addition allows for higher deposition rates.

## Acknowledgements

This work is supported by the Swedish Research Council (Grant no. 621-2010-4011), the Swedish Energy Agency (Grant no. 32096-1) and the STandUP for energy strategic research framework. The authors would also like to thank Dr. Örjan Wallin from the Division of Solid state electronics, Department of Engineering Sciences, Uppsala University for help with deposition of SiO$_2$ by TEOS.

## References

1. J.Isberg, J.Hammersberg, E.Johansson, T.Wikstrom, D.J.Twitchen, A.J.Whitehead, S.E.Coe, and G.A.Scarsbrook, *Science* **297**, 1670 (2002).

2. H.Taniuchi, H.Umezawa, T.Arima, M.Tachiki, H. Kawarada, *IEEE Electron Device Lett.*, **22** , 390, (2001).
3. K.Hirama, S.Miyamoto, H.Matsudaira, K.Yamada, H.Kawarada, T.Chikyo, H. Koinuma, K. Hasegawa, and H. Umezawa, *Appl. Phys. Lett*. **88**, 112117 (2006).
4. K.Oyama, S.G. Ri, H.Kato, M.Ogura, T.Makino, D.Takeuchi, N.Tokuda H.Okushi, and S.Yamasaki, *Appl. Phys. Lett*, **94**, 152109 (2009).
5. S.J. Rashid, A. Tajani, L. Coulbeck, M. Brezeanu, A. Garraway, T. Butler, N.L. Rupesinghe, D.J. Twitchen, G.A.J. Amaratunga, F. Udrea, P. Taylor, M. Dixon, J. Isberg, *Diam. and Relat. Mater*. **15**, (2–3), 317-323 , (2006).
6. K.K.Kovi, S.Majdi, M.Gabrysch, I. Friel, R.Balmer and J.Isberg, *MRS fall meeting 2010 proceedings*, **1282**, (2011).
7. R.S.Balmer, I.Friel, S.M.Woollard, C.J.H.Wort, G.A.Scarsbrook, S.E.Coe, H.El-Hajj, A.Kaiser, A.Denisenko, E.Kohn, and J.Isberg, *Phil. Trans. Roy. Soc. A*: **366,** 251-265, (2008).
8. H.Pernegger, S.Roe, P.Weilhammer, V.Eremin, H.Frais-Kolbl, E Griesmayer, H.Kagan, S.Schnetzer, R.Stone, W.Trischuk, D.J.Twitchen, A.Whitehead, *J.Appl.Phys*. **97**, 1 (2005).
9. C.E.Nebel, J.Muenz, M.Stutzmann, R.Zachai, *Phys. Rev. B: Condens. Matter* **55**,(15), 9786 (1997).
10. F.Fujita, A.Homma, Y.Oshiki, J.H.Kaneko, K.Tsuji, K.Meguro, Y.Yamamoto, T.Imai, T.Teraji, T.Sawamura, M.Furusaka, *Diamond Relat. Mater.* **14** (11–12),1992 (2005).
11. M.Nesladek, A. Bogdan, W.Deferme, N.Tranchant, P.Bergonzo, *Diam. Relat. Mater*. **17** (7–10), 1235 (2008).
12. J.Isberg, J.Hammersberg, D.J.Twitchen, A.Whitehead, *Diam. Relat. Mater*. **13**, (2), 320, (2004).
13. M.Pomorski, E.Berdermann, M.Ciobanu, A.Martemyianov, P.Moritz, M.Rebisz, B.Marczewska, *Phys. Status Solidi A* **202**, (11), 2199, (2005).
14. J.Isberg, M.Gabrysch, A.Tajani, D.J.Twitchen, *Semicond. Sci. Technol.* **21** (8),1193 (2006).
15. J.Isberg, A.Lindblom, A.Tajani, D.J.Twitchen, *Phys. Status Solidi A* **202** (11), 2194, (2005).
16. M.Pomorski, E.Berdermann, A.Caragheorgheopol, M.Ciobanu, M. Kis, A.Martemiyanov, C. Nebel, P. Moritz, *Phys. Status Solidi A* **203** (12), 3152, (2006).
17. M. Pomorski, E. Berdermann, W. de Boer, A. Furgeri, C. Sander, J. Morse, *Diamond Rel.Mater.* **16** (4–7), 1066, (2007).
18. W.Deferme, A.Bogdan, G.Bogdan, K.Haenen, W.DeCeuninck, M.Nesladek, *Phys. Stat. Sol. A* **204** (9), 3017, (2007).
19. Y.Oshiki, J.H.Kaneko, F.Fujita, K.Hayashi, K.Meguro, A.Homma, S.Kawamura, Y.Yokota, Y.Yamamoto, K.Kobashi, T.Imai, T.Sawamura, M.Furusaka, *Diam.Rel.Mater.* **15** (10), 1508, (2006).
20. Y.Oshiki, J.H.Kaneko, F.Fujita, A.Homma, H.Watanabe, K.Meguro, Y.Yamamoto, T.Imai, K. Sato, K. Tsuji, S. Kawamura, M. Furusaka, *Diam. Rel.Mater*.**17** (4–5), 833, (2008).
21. J.Isberg, S.Majdi, M.Gabrysch, I.Friel, R.S.Balmer, *Diam. Relat. Mater.* **18,** 1163-1166 (2009).
22. G.Ottaviani, C.Canali, F.Nava, and J.W.Mayer, *J. Appl. Phys*. **44**, 2917,(1973).
23. L.S.Pan, S.Han, D.R.Kania, S.Zhao, K.K.Gan, H.Kagan, R. Kass, R.Malchow, F.Morrow, W.F.Palmer, C.White, S.K.Kim, F.Sannes, S.Schnetzer, R.Stone, G.B.Thomson, Y.Sugimoto, A.Fry, S.Kanda, S.Olsen, M.Franklin, J.W.Ager, and P.Pianetta, *J. Appl. Phys*. **74**, 1086,(1993).
24. M.Gabrysch, E.Marklund, J.Hajdu, D.J.Twitchen, J.Rudati, A.M.Lindenberg, C.Caleman, R.W.Falcone, T.Tschentscher, K.Moffat, P.H.Bucksbaum, J.Als-Nielsen, A.J.Nelson,

D.P.Siddons, P.J.Emma, P. Krejcik, H.Schlarb, J.Arthur, S.Brennan, J.Hastings, and J.Isberg, *J. Appl. Phys*. **103**, 064909, (2008).
25. A.Denisenko and E.Kohn, *Diam. Relat. Mater*. **14**, 491-498,(2005).
26. J.R. Haynes, W.Shockley, *Phys. Rev.* **81**, 835,(1951).
27. J.Isberg, M.Gabrysch, S.Majdi, K.K.Kovi, D.J.Twitchen, *Solid State Sciences* **13** (5), 1065-1067, (2011).

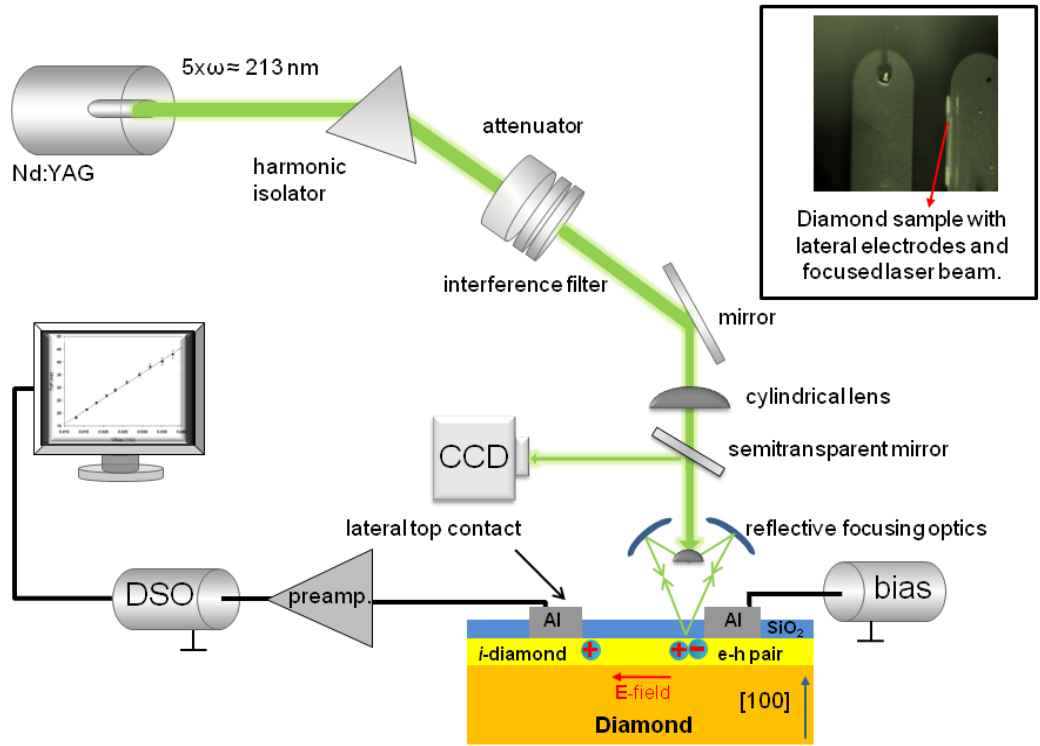

Fig 1: Schematics of the ToF setup. The sample is illuminated with 3 ns (FWHM) 213 nm UV light from a quintupled Nd-YAG laser. Holes drift from the right contact towards the left contact and the resulting current is measured using a broadband amplifier; with the laser illuminating a line close to one of the contacts. A bond wire can also be seen on the left contact (*top right*).

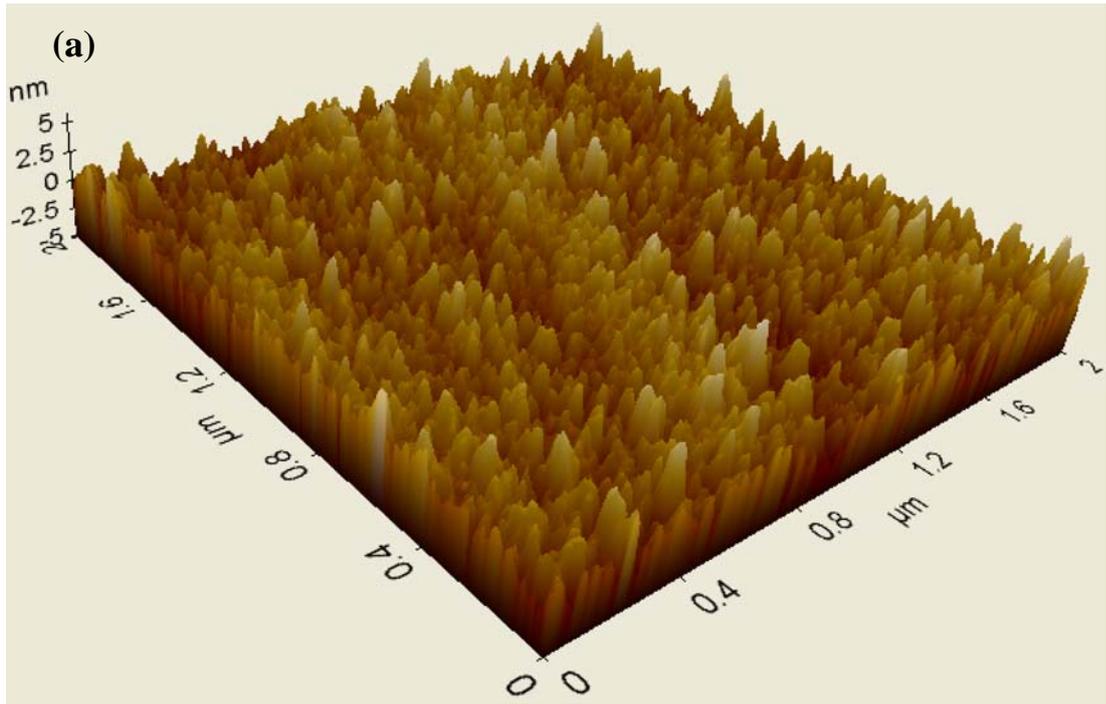
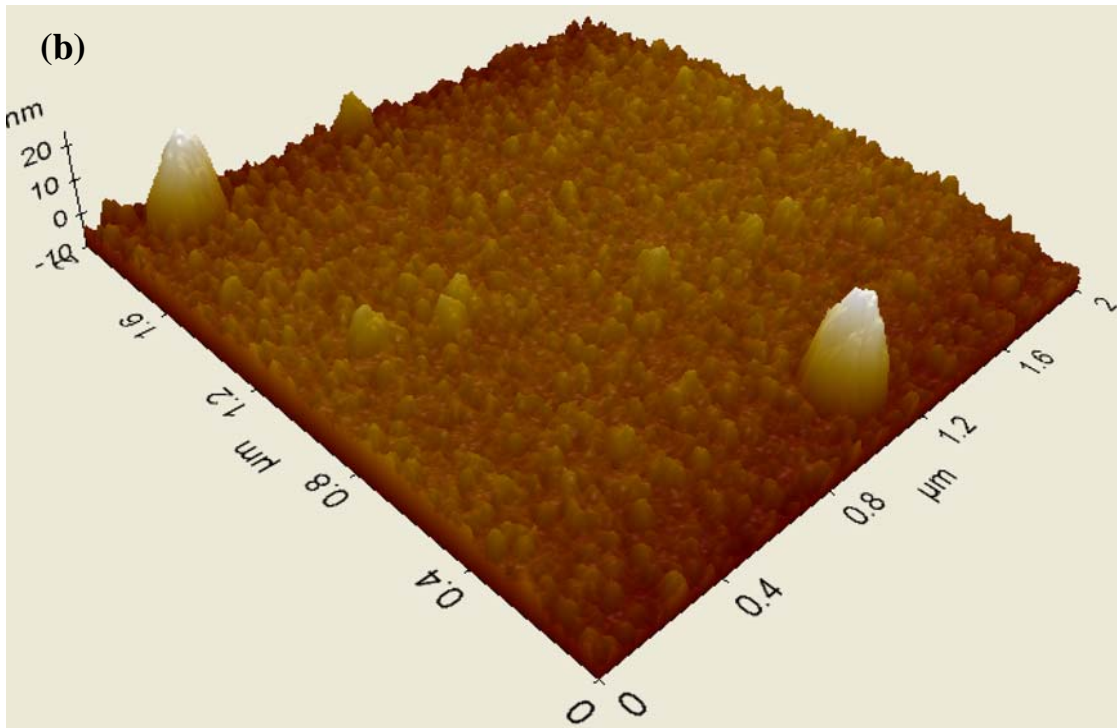

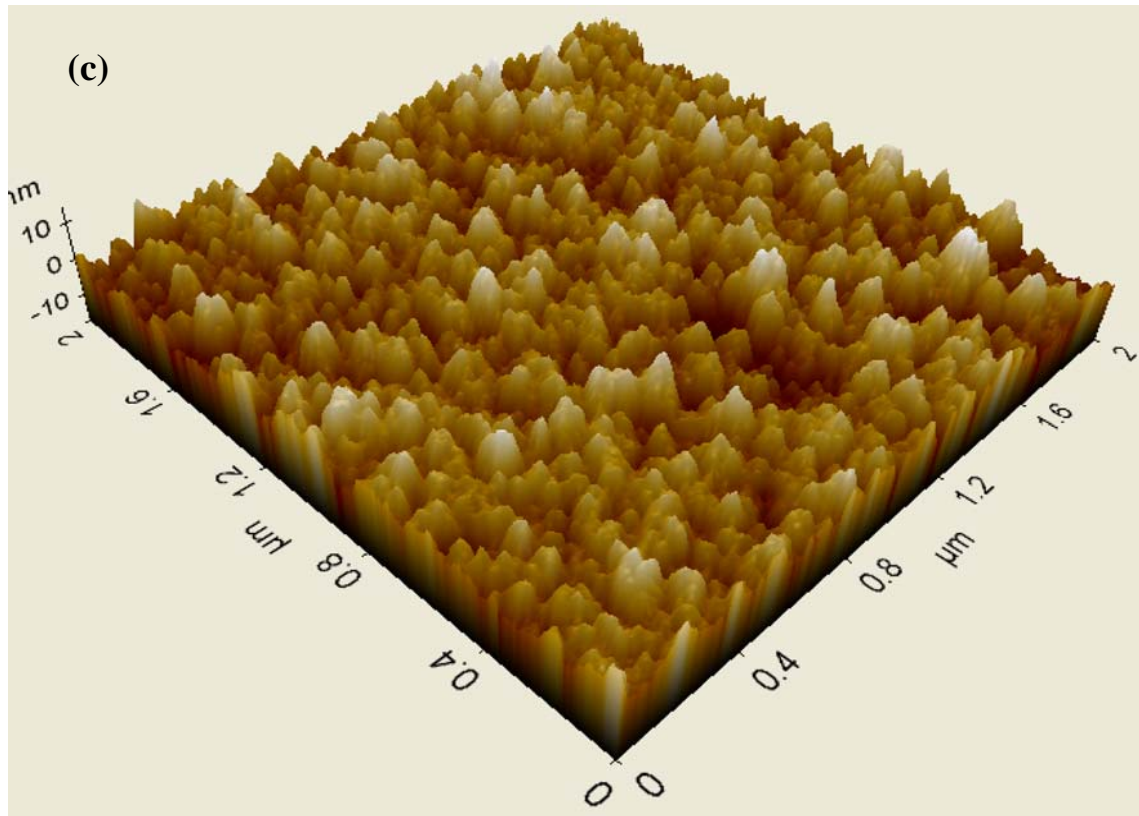

Fig 2: AFM images of (a) LPCVD-TEOS, (b) PVD and (c) PECVD deposited SiO$_2$ layers.

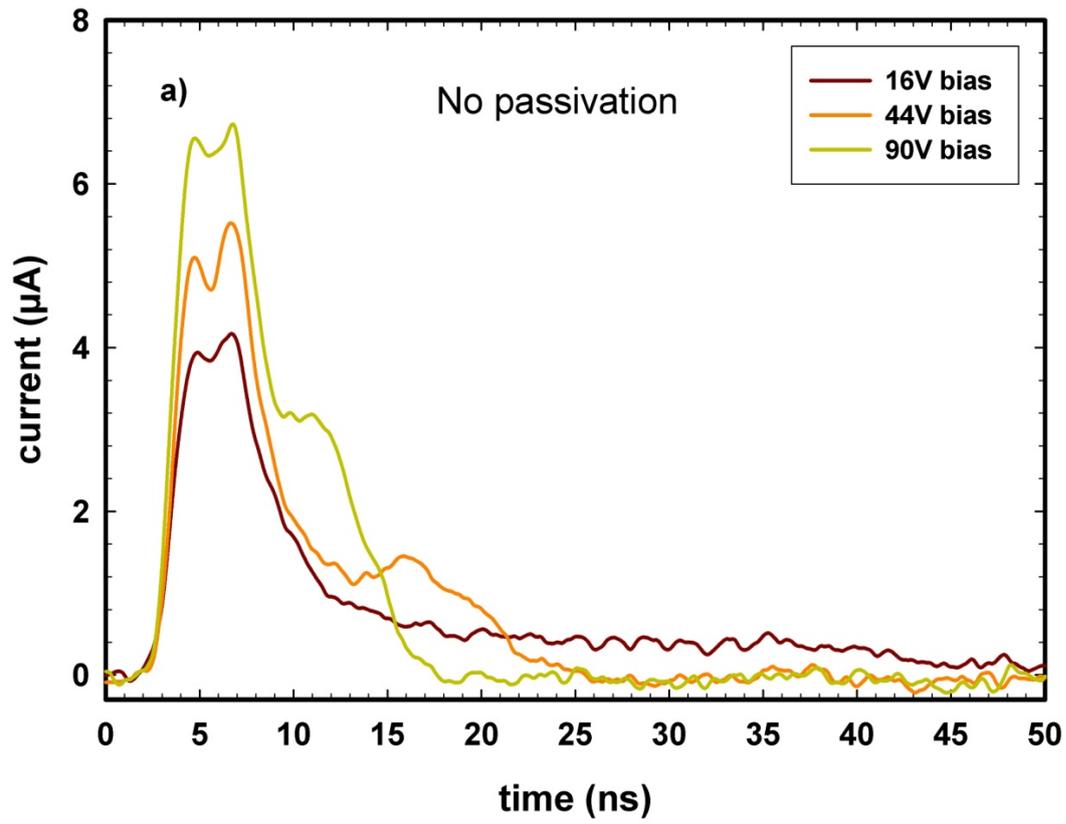
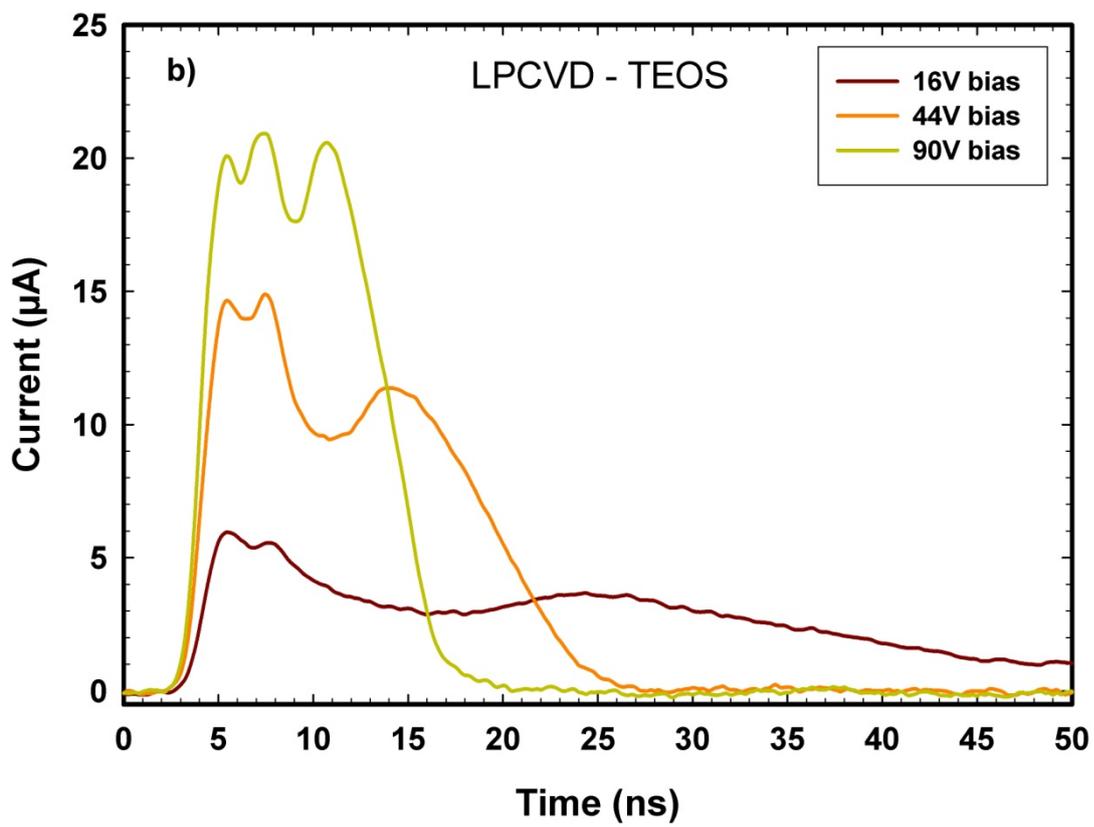

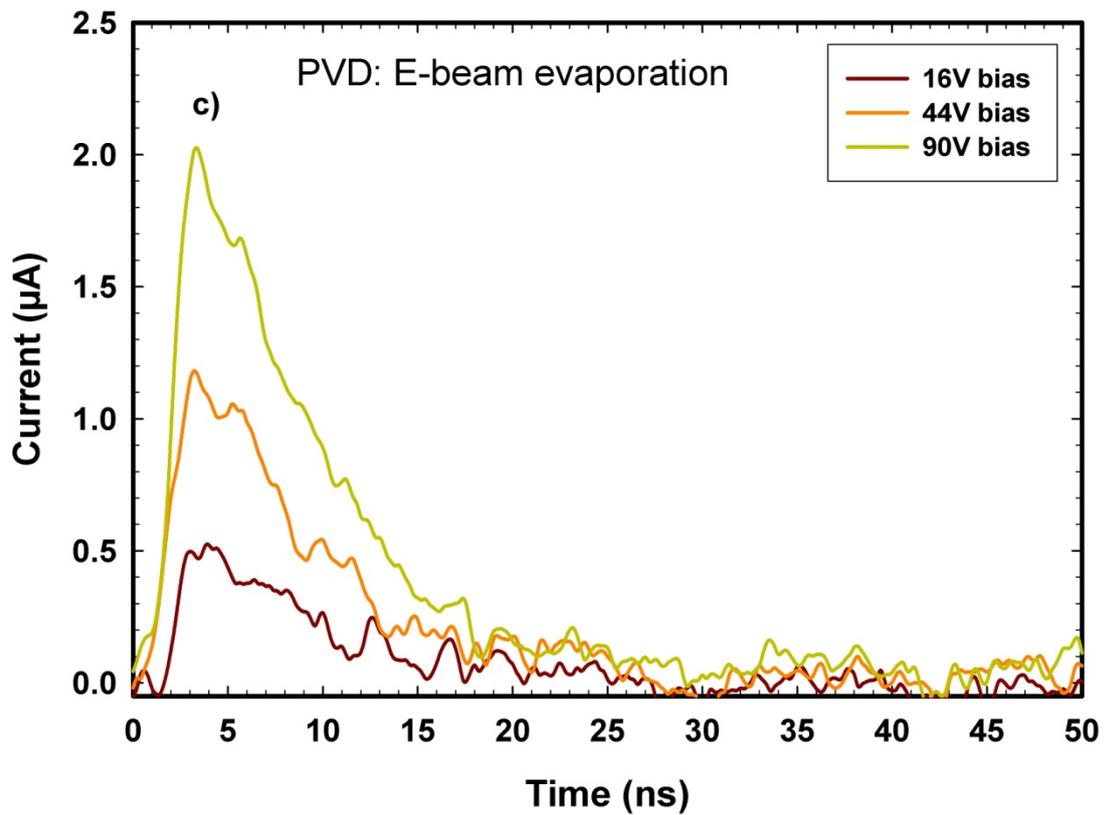

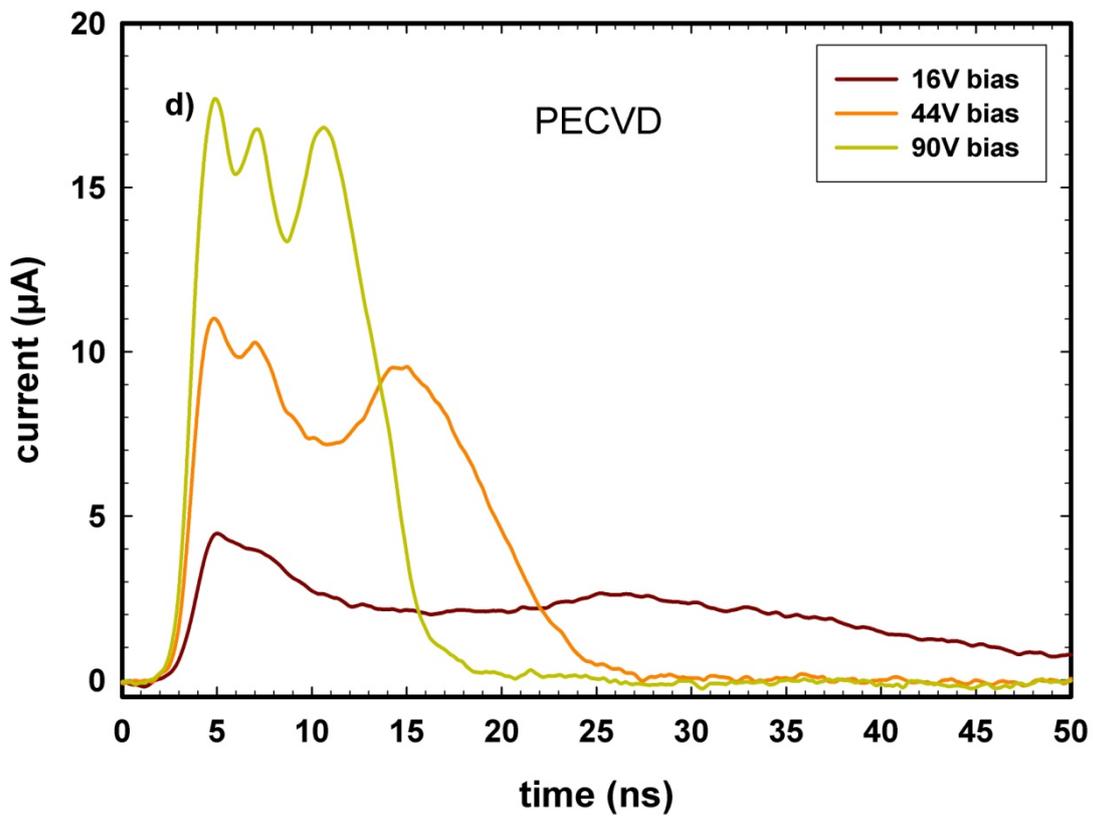

Fig 3: Comparison of current traces with different surface passivation layers at different bias voltages indicated by different colours. (a) No passivation, (b) LPCVD-TEOS, (c) PVD (d)

PECVD. The PVD method shows poor signal strength and a dispersive transient. PECVD & LPCVD SiO$_2$ layers display higher charge collection compared to PVD-SiO$_2$. In addition the arrival of the holes at the receiving contact can clearly be seen in these traces, indicating a much reduced rate of hole trapping at the interface.

| Type of SiO$_2$ | Hole mobility (cm$^2$/Vs) | Thickness of SiO$_2$ (nm) | Surface roughness (nm) (from AFM) |
|---|---|---|---|
| LPCVD | 1410 ± 30 | 360 | 2 |
| PECVD | 1360 ± 25 | 380 | 5 |
| PVD | - | 125 | 10-12 |
| No passivation | 1300 ± 20 | - | 1 |

Table 1: The obtained surface roughness measured from AFM and the calculated hole mobilities from the Time-of-Flight measurements on the same sample with SiO$_2$ deposited by different methods.